\documentclass[12pt,nohyper,letterpaper]{JHEP3}         
\usepackage{amssymb,amsfonts}

\usepackage{epsfig}

\def\be{\begin{eqnarray}}
\def\ee{\end{eqnarray}}
\newcommand{\nn}{\nonumber}
\newcommand\para{\paragraph{}}
\newcommand{\ft}[2]{{\textstyle\frac{#1}{#2}}}
\newcommand{\eqn}[1]{(\ref{#1})}

\def\Dslash{\,\,{\raise.15ex\hbox{/}\mkern-12mu D}}
\def\Dbarslash{\,\,{\raise.15ex\hbox{/}\mkern-12mu {\bar D}}}
\def\delslash{\,\,{\raise.15ex\hbox{/}\mkern-9mu \partial}}
\def\delbarslash{\,\,{\raise.15ex\hbox{/}\mkern-9mu {\bar\partial}}}
\def\pslash{\,\,{\raise.15ex\hbox{/}\mkern-9mu p}}
\def\calDslash{\,\,{\raise.15ex\hbox{/}\mkern-12mu {\cal D}}}

\newcommand{\sign}{{\rm sign}}

\newcommand{\nin}{\,{\raise.15ex\hbox{/}\mkern-12mu \!\in}}

\newcommand{\Tr}{{\rm Tr}}

\def\lae{\mathrel{\mathop{\smash{\lower .5 ex \hbox{$\stackrel<\sim$}}}}}
\def\lae{\mathrel{\mathop{\smash{\lower .5 ex \hbox{$\stackrel>\sim$}}}}}

\title{The Dynamics of Chern-Simons Vortices}
\author{Benjamin Collie and David Tong \\
Department of Applied Mathematics and Theoretical Physics, \\
University of Cambridge, UK\\
{\tt b.p.collie, d.tong@damtp.cam.ac.uk}}

\abstract{We study vortex dynamics in three-dimensional theories
with Chern-Simons interactions. The dynamics is governed by motion
on the moduli space ${\cal M}$ in the presence of a magnetic
field. For Abelian vortices, the magnetic field is shown to be the Ricci form
over ${\cal M}$; for non-Abelian vortices, it is the first Chern
character of a suitable index bundle. We derive these results by
integrating out massive fermions and following the fate of their
zero modes.}

\begin{document}
\pagestyle{plain} \setcounter{page}{1}
\newcounter{bean}
\baselineskip16pt

\section{Introduction}

The moduli space approximation provides an elegant description of
the low-energy behaviour of solitons \cite{manton}. Information
about soliton interactions is packaged in a simple geometric form
which has proven useful in extracting both the classical and
quantum dynamics of the system. In this paper we use the moduli
space approximation to study the motion of vortices in the
presence of Chern-Simons interactions \cite{cs}.

\para
For vortices in the Abelian-Higgs model in $d=2+1$ dimensions, a
moduli space ${\cal M}$ of solutions exists only when the
potential is tuned to critical coupling, meaning that the theory
lies on the borderline between Type I and Type II
superconductivity. For $k$ vortices, the moduli space has
dimension ${\rm dim}({\cal M})= 2k$, with the coordinates $X^a$,
$a=1,\ldots, 2k$, on ${\cal M}$ corresponding to the positions of
the vortices on the plane \cite{erick,taubes}. At low-energies,
the scattering of vortices can be described as geodesic motion on
${\cal M}$ with respect to a metric $g_{ab}$,
\be L_{\rm vortex} = \ft12
g_{ab}(X)\dot{X}^a\dot{X}^b\label{samols}\ee
Although the metric $g_{ab}$ is not known explicitly for $k\geq
2$, its properties have been well studied
\cite{samols,mansp,manchen}. Most notably, $g_{ab}$ is K\"ahler.

\para
One can  ask how the dynamics of the vortices is affected by the
addition of a Chern-Simons interaction
\cite{csdyn,kimyeongold,kimyeong}. On general grounds, one expects
the low-energy dynamics of vortices to be governed by geodesic
motion on ${\cal M}$, now in the presence of a magnetic field
${\cal F}\in \Omega^2({\cal M})$. Locally we may write ${\cal
F}=d{\cal A}$ and the Lagrangian takes the form,
\be L_{\rm vortex} = \ft12 \tilde{g}_{ab}(X)\dot{X}^a\dot{X}^b -
\kappa {\cal A}_a(X)\dot{X}^a\label{right}\ee
where $\kappa$ is the coefficient of the Chern-Simons term in
three-dimensions. Working perturbatively in $\kappa$, Kim and Lee
found that to leading order $\tilde{g}_{ab}=g_{ab}$, while an
expression for ${\cal A}$ was given in terms of the profile
functions of the vortices \cite{kimyeong}. However, the geometric
meaning of ${\cal A}$ has remained mysterious. Here we remedy
this. We show that ${\cal F}$ is the Ricci form on ${\cal M}$.

\para
We further study the dynamics of non-Abelian $U(N)$ vortices
introduced in \cite{vib,auzzi} in the presence of Chern-Simons
interactions. In this case the moduli space has dimension
$\dim({\cal M})=2kN$ and the dynamics is again given by
\eqn{right}. We show that ${\cal F}$ is the first Chern character
of a particular index bundle over ${\cal M}$.

\para
The technique we use to derive these results is simple yet
indirect, and can be viewed as an application of the
Goldstone-Wilczek method \cite{gw,cw}. We make use of the
well-known fact that the Chern-Simons terms can be induced by
integrating out heavy fermions in three dimensions
\cite{redlich,agw}. We follow the fate of these fermions from the
perspective of the vortices. The fermi zero modes live in an index
bundle over ${\cal M}$ and we show that, as their mass becomes
large, they may be integrated out to reproduce the result
\eqn{right}. It is then simple to show that there is no further
contribution from non-zero modes. We recently employed this method
to derive the dynamics of instantons in five-dimensional Yang-Mills
Chern-Simons theories \cite{5dcs}.

\para
The plan of the paper is as follows: in Section 2 we introduce the
model of interest and describe its vortex solutions. It is a
$U(N)$ Yang-Mills theory, with Chern-Simons interactions, coupled
to matter fields. The Lagrangian admits ${\cal N}=2$ supersymmetry
and the vortices are BPS. In Section 3 we present our main
results, analyzing the impact on the vortex dynamics as fermions
are introduced, made heavy, and finally integrated out. Section 4
is devoted to two examples. In the first example, we study the
qualitative dynamics of two Abelian vortices and describe the
bound orbits. We also show that our technique correctly reproduces
the fractional statistics of Abelian vortices. The second example
concerns a single vortex in the $U(N)$ theory for which the moduli
space is ${\bf CP}^{N-1}$ and the appropriate magnetic field
${\cal F}$ is proportional to $\Omega$, the K\"ahler form. We also
show how to reproduce this magnetic field from a direct study of
the vortex equations in the moduli space approximation.

\section{The Vortex Equations}

The literature contains a veritable smorgasbord of
Chern-Simons models which admit vortex solutions. These include
Abelian theories with \cite{llm} and without \cite{hkp,jw,jlw} a
Maxwell term, non-Abelian theories \cite{klee,schap}, and theories
with non-relativistic kinetic terms for the matter fields
\cite{zhk,pij,jip,nonrel}. The properties of many of these models are
summarized in the excellent review \cite{dunne}.

\para
Our interest in this paper lies in a $U(N)$
Yang-Mills-Chern-Simons theory coupled to a real adjoint scalar
$\phi$ and $N_f$ scalars $q_i$, $i=1,\ldots,N_f$, each of which
transforms in the fundamental representation of the gauge group.
With suitable fermion content, the theory enjoys ${\cal N}=2$
supersymmetry (i.e. 4 four supercharges) which dictates the form
of the bosonic interactions:
\be {\cal L} &=& -\frac{1}{2e^2}\Tr\,F_{\mu\nu}F^{\mu\nu} -
\frac{\kappa}{4\pi}\Tr\,\epsilon^{\mu\nu\rho}(A_\mu \partial_\nu
A_\rho - \frac{2i}{3}A_\mu A_\nu A_\rho) +
\frac{1}{e^2}\Tr\,({\cal D}_\mu\phi)^2 \nn\\ && + |{\cal D}_\mu
q_i|^2 - q_i{}^{\!\dagger}\phi^2 q_i -
\frac{e^2}{4}\Tr(q_iq_i{}^{\!\dagger} -
\kappa\phi/2\pi-v^2)^2\label{lag}\ee
Notice that we have not considered separate Chern-Simons
coefficients for the $U(1)$ and $SU(N)$ parts of the gauge group,
but instead taken a specific combination in which they are
packaged together in $U(N)$. For $N\geq 2$, invariance of the
partition function under large gauge transformation requires that
$\kappa \in {\bf Z}$. For the Abelian theory, there is no such
constraint.

\para
To make contact with the other models on the market, it is
instructive to consider various limits of this Lagrangian.
\begin{itemize}
\item For $U(1)$ gauge group, the Lagrangian reduces to the
Maxwell-Chern-Simons-Higgs theory introduced in \cite{llm}.
\item When $\kappa=0$, the Lagrangian reduces to Yang-Mills theory
coupled to a number of fundamental scalar fields. This theory is
known to admit non-Abelian vortices, first introduced in
\cite{vib,auzzi} and since studied in some detail. (See, for
example, \cite{tasi,moduli,sy} for reviews). We will make much use of this limit.
\item When $e^2\rightarrow\infty$, the Yang-Mills term vanishes,
and the scalar field $\phi$ becomes auxiliary. Integrating out
$\phi$ reproduces the Chern-Simons-Higgs theory with sixth order
scalar potential, first introduced in the Abelian case in
\cite{hkp,jw,jlw}, and studied more recently in the non-Abelian
case in \cite{schap}. \end{itemize}

\para
Two important ground states of the theory are the unbroken phase and the Higgs phase.
The gauge symmetry is unbroken when the scalar fields take
the vacuum expectation values
\be \mbox{Unbroken Phase}:\ \ \ \ \ \phi^a_{\ b} = -\frac{2\pi
v^2}{\kappa}\delta^a_{\ b} \ \ \ ,\ \ \ q_i=0\label{unbroken}\ee
where $a,b=1,\ldots,N$ is the colour index. This state exists
regardless of the number $N_f$ of fundamental flavours. In
contrast, a ground state with fully broken gauge symmetry only
exists when $N_f \geq N$ and the rank $N_f$ term
$q_iq_i{}^{\!\dagger}$ in the potential can successfully cancel
the rank $N$ term $v^2$ (which comes with an implicit $N\times N$
unit matrix). For simplicity, in what follows we choose $N_f=N$.
There is then a unique ground state with fully broken gauge symmetry
given by
\be \mbox{Higgs Phase}:\ \ \ \ \ \ \ \ \ \phi =0 \ \ \ \ ,\ \ \
q_i^{\ a} = v\delta_i^{\ a}\label{HiggsPhase}\ee
In this vacuum, both the $U(N)$
gauge symmetry and the $SU(N)$ flavour symmetry which rotates the
Higgs fields $q_i$ are spontaneously broken. However, the diagonal
of the two survives: $U(N)_{\rm gauge}\times SU(N)_{\rm flavour}
\rightarrow SU(N)_{\rm diag}$.
The theory also has several ground states with partly broken gauge symmetry.  
For each such state, the vacuum expectation values of the fields have some 
diagonal entries equal to those in \eqn{unbroken} and the rest equal to those 
in \eqn{HiggsPhase}.  We will not consider these partly broken phases further.

\para
In the Higgs phase, the model admits topologically stable BPS
vortices. First order equations of motion may be derived using the
standard Bogomolnyi trick, and read
%
%
%
%
\be B = \frac{e^2}{2} (q_i q_i^\dagger - \kappa\phi/2\pi - v^2) \
\ ,\
\ \ {\cal D}_zq_i \equiv {\cal D}_1q_i-i{\cal D}_2q_i=0 \label{bog1}\\
E_\alpha + {\cal D}_\alpha\phi =0 \ \ \ ,\ \ \ {\cal D}_0\phi=0\ \
\ ,\ \ \ {\cal D}_0q_i+i\phi q_i=0\label{bog2}\ee
Here $B=F_{12}$ and $E_\alpha = F_{0\alpha}$.  Note however that,
in contrast to vortices in $\kappa =0$ theories, it is not enough
to solve these first order equations alone: we must also solve
Gauss' law. This is most simply written in static gauge
$\partial_0=0$. Then the three equations in \eqn{bog2} may all be
solved by setting $A_0=\phi$, which is determined by Gauss' law
\be 2{\cal D}^2\phi +\frac{\kappa}{2\pi} e^2 B -
e^2\{\phi,q_iq_i^\dagger\}=0\label{gauss}\ee
Note that the presence of the Chern-Simons coupling ensures that
$\phi$ is sourced at the core of the vortex where $B\neq 0$. The
fact that the first order vortex equations \eqn{bog1} must be
supplemented by the second order equation \eqn{gauss} is what makes the study
of vortex dynamics somewhat more of a technical challenge in the
presence of a Chern-Simons interaction.

\para
Configurations satisfying \eqn{bog1} and \eqn{gauss} have energy,
\be E = -v^2 \int d^2x \ \Tr\,B = 2\pi v^2 k\ee
where $k\in {\bf Z}^+$ is the topological charge of the vortex. It
is expected that equations \eqn{bog1} and \eqn{gauss} enjoy a
moduli space of solutions of dimension $\dim({\cal M})=2kN$. This
is suggested by the counting of zero modes using index theorems
\cite{erick,hkp,vib}. However, to our knowledge the only rigorous
proof of this statement when $\kappa\neq 0$ holds in the
Abelian theory in the limit $e^2\rightarrow \infty$ \cite{wang}.
To some extent, the method we propose in the next section
circumvents this issue since our starting point will be the theory
with $\kappa =0$ where the existence of a moduli space has been
rigorously proven \cite{taubes}.

\para
Our theory has ${\cal N}=2$ supersymmetry. Yet so far we have not
mentioned the fermions. They consist of a single Dirac fermion
$\lambda$ in the adjoint representation of the gauge group (this
is the superpartner of $A_\mu$ and $\phi$) together with $N_f$
Dirac fermions $\psi_i$ in the fundamental representation  (the
superpartners of $q_i$). In the background of the vortex, these
fermions carry zero modes. These zero modes will not be the focus of our
discussion in the next section, although one should remember that
they are present. Instead we will be interested in the zero modes
of some extra, supplementary, fermions that we now introduce.

\section{Integrating Out Fermions}

Our strategy in this section is to replace the Chern-Simons
interactions in the bosonic Lagrangian \eqn{lag} with something
that we understand better, namely fermions. To this end, we start
with the theory without Chern-Simons interactions by setting
$\kappa=0$ in \eqn{lag}. We now introduce $\tilde{N}$ chiral
multiplets $\tilde{Q}$, each transforming in the anti-fundamental
representation of the $U(N)$ gauge group. Each of these chiral
multiplets may be given a mass $m$ consistent with
supersymmetry\footnote{This mass term is not possible in $d=3+1$
dimensions, where it would break Lorentz invariance. It is allowed
in $d=2+1$, and was called a ``real mass" in \cite{ahiss} to
distinguish it from the more familiar complex mass that appears in
the superpotential. For the present purposes, the important point
is its effect on the fermions which is shown in the Dirac equation
\eqn{dirac}.}. In the limit $m\rightarrow \infty$, the chiral
multiplets may be happily integrated out. All of their effects
decouple, except for a remnant $U(N)$ Chern-Simons term, with
coefficient \cite{redlich,agw}
\be \kappa = - \frac{\tilde{N}}{2}\,{\rm sign}(m)\label{nkappa}\ee
Importantly, the $\kappa\phi$ term in the potential in \eqn{lag}
is the supersymmetric partner of the Chern-Simons term.
Supersymmetry is unbroken in this model (the Witten index is
non-vanishing), so when we integrate out the chiral multiplets $\tilde{Q}$,
the $\kappa\phi$ term must be generated together with the Chern-Simons term.
In fact, there is one further, related,
effect that is important: the scalar vev $v^2$ (which is a
Fayet-Iliopoulos parameter in the language of supersymmetry) picks
up a finite renormalization \cite{ahiss,memirror}:
\be v^2 \rightarrow v^2_{\rm eff} = v^2 + \frac{m \kappa}{2\pi} =
v^2 - \frac{\tilde{N}|m|}{4\pi}\label{veff}\ee
Notice that for suitably large $|m|$, we have $v^2 <0$, and the
theory exits the Higgs phase where the vortices live. If we wish
to stay in the Higgs phase, and keep the vortex mass fixed, we
must scale $v^2$ so that $v^2_{\rm eff}$ remains constant as
$m\rightarrow\infty$. If we perform such a scaling, we conclude
that the Chern-Simons theory \eqn{lag} is equivalent to the
Yang-Mills theory coupled to $\tilde{N}=2\kappa$ supplementary
massive chiral multiplets $\tilde{Q}$ in the limit $m\rightarrow
\pm \infty$.

\subsection{The Index Bundle of Fermi Zero Modes}

Let us now follow the effect of this procedure on the vortex
dynamics, focussing firstly on a  Dirac fermion $\tilde{\psi}$ in
one of the chiral multiplets $\tilde{Q}$. The Dirac equation is
given by,
%
%
\be i\Dslash \tilde{\psi} - \tilde{\psi}\phi =
m\tilde{\psi}\label{dirac}\ee
We are interested in the solutions to this equation in the
background of the vortex. Since we are working in the theory with
$\kappa=0$, the bosonic fields of the vortex are solutions to
\be B=\frac{e^2}{2}(q_iq_i^\dagger - v^2)\ \ \ ,\ \ \ {\cal D}_z
q_i =0\ \ \ , \ \ \ A_0=\phi=0\label{veq}\ee
Of the full spectrum of solutions to the Dirac equation
\eqn{dirac} in the background of the vortex, only the zero modes
will prove important. We discuss these first, returning to the
non-zero modes shortly. We work with the basis of gamma matrices
$\gamma^\mu = (\sigma^3,i\sigma^2,-i\sigma^1)$. The zero modes
then take the form \cite{jr}
\be \tilde{\psi}(t,x_\alpha) = e^{-imt}
\left(\begin{array}{c} \tilde{\psi}_-(x_\alpha)  \\
0\end{array}\right)\ \ \ {\rm or}\ \ \ \ \tilde{\psi}(t,x_\alpha)
= e^{+imt}
\left(\begin{array}{c} 0  \\
\tilde{\psi}_+(x_\alpha)\end{array}\right) \ee
where ${\cal D}_{\bar{z}}\tilde{\psi}_- = {\cal
D}_z\tilde{\psi}_+=0$. Standard index theorems state that the
equation ${\cal D}_{\bar{z}}\tilde{\psi}_-=0$ has $k$ solutions in
the background of the vortex, while ${\cal D}_z\tilde{\psi}_+=0$
has none\footnote{Recall that $\tilde{\psi}$ transforms in the
$\bar{\bf N}$ representation, while $q_i$ transforms in the ${\bf
N}$ representation -- this is responsible for the fact that ${\cal
D}_{\bar{z}}$ carries the zero modes in the background ${\cal
D}_zq_i=0$. More details on these fermi zero modes in the context
of vortex strings in related four-dimensional theories can be
found in \cite{het}.}. For example, in the Abelian case this
follows from the fact that there is no holomorphic line bundle of
negative degree.

\para
The space of zero modes of the Dirac equation defines a bundle
over the vortex moduli space ${\cal M}$, with fibre ${\bf C}^k$.
This is commonly referred to as the index bundle \cite{mansch,wy}.
As we move in moduli space by adiabatically changing the
background vortex configuration, the fermi zero modes undergo a
holonomy described by a Hermitian $u(k)$ connection $\omega$ over ${\cal
M}$. We denote the Grassmann-valued coordinates of the fibre as
$\xi^l$, $l=1,\ldots, k$. The low-energy dynamics of the vortex
should now be augmented to include these zero modes, described by
the kinetic terms\footnote{\label{footnote}There is an important
caveat here: the zero modes under discussion are non-normalizable;
they have a long-range $1/r$ tail, causing them to suffer from an
infra-red logarithmic divergence. In the context of
four-dimensional theories, there are several examples where
ignoring this fact, and treating these modes with kinetic terms of
the form \eqn{goodgood}, leads to quantitatively and qualitatively
correct physics \cite{vstring,qhet}. This approach has been
criticized in \cite{syok}. For the time being, we proceed by
ignoring this issue. However, in Section \ref{abstring} we will
present a slightly more involved construction that yields the same
answer, but doesn't suffer from this technical problem.}
\be L = \bar{\xi}^l(iD_t-m)\xi^l\label{goodgood}\ee
where the covariant derivative is defined by
\be D_t\xi^l = \partial_t\xi^l + i(\omega_a)^l_{\ m}\dot{X}^a\,
\xi^m\ee
Let us pause briefly to discuss how one should quantize these zero
modes. As usual, each complex fermionic zero mode gives rise to two states
--- occupied and unoccupied --- whose energy differs by $m$.
However, the question of the absolute ground state energy requires
us to resolve the usual ordering ambiguities. Comparison with the
renormalization of $v^2$ given in \eqn{veff} shows that a single
fermi zero mode should cause the mass of the vortex to shift by
$M_{\rm vortex} \rightarrow M_{\rm vortex} - |m|/2$. This strongly
suggests that we should take the ground state of the fermi zero
modes to have energy $-|m|/2$, and the excited state to have
energy $+|m|/2$. It would be interesting to understand better why
this choice of ordering is forced upon us.

\subsection{Integrating out the Index Bundle}

Throughout this discussion, we have been referring to the relevant
solutions of the Dirac equation as ``zero modes". This is a slight
misnomer because, as is clear from \eqn{goodgood}, they are
excited at a cost of energy equal to $|m|$. They become true zero
modes only in the $m\rightarrow 0$ limit which is, of course, to
be expected since they arose from fermions with mass $m$. However,
we are interested in the opposite limit $m\rightarrow \infty$. In
this limit, the effect of the fermi zero modes {\it almost}
decouples. They do not correct the metric $g_{ab}$. However, as we
now show, they do give rise to new Chern-Simons terms for the
moduli space dynamics.

\para
Integrating out the fermion $\xi$ in the path integral leads to
the ratio of determinants
\be \det\left(\frac{iD_t-m}{i\partial_t-m}\right)\ee
We can compute this ratio using standard methods. We work with
compact Euclidean time $\tau=it$, with periodicity
$\tau\in[0,\beta)$.  We look for the eigenvalues $\lambda$ of the
operator
\be \left(-\partial_\tau - i\omega - m\right) \chi = \lambda\chi\ee
where $\omega=\omega_a\,\partial_\tau{X}^a$. The eigenfunctions
are subject to periodic boundary conditions $\chi(0)=\chi(\beta)$.
Solutions are given by the usual time-ordered product
\be \chi=e^{-(m+\lambda)\tau}\,V(\tau)\chi\ \ \ \ {\rm with}\ \ \
V(\tau) =
T\,\exp\left(-i\int_0^{\tau}d\tau^\prime\omega(\tau^\prime)\right)
\in U(k)\ee
Let us denote the eigenvalues of $V(\beta)$ as $e^{v_l}$,
$l=1,\ldots,k$. Then the periodicity requirement
$\chi(0)=\chi(\beta)$ means that the eigenvalues $\lambda$ are
given by
\be \lambda = \frac{2\pi i n+v_l}{\beta} -m\ \ \ \ \ \ \ n\in {\bf
Z},\ \  l=1,\ldots,k\ee
>From this we compute the ratio of determinants
\be \det\left(\frac{D_\tau+m}{\partial_\tau+m}\right) &=&\ \
\prod_{l=1}^k\prod_{n\in {\bf Z}}\left(\frac{2\pi i n/\beta +
v_l/\beta - m}{2\pi i n/\beta - m}\right) \nn\\ &=&\ \
\prod_{l=1}^k
\left(1-\frac{v_l}{m\beta}\right)\,\left(\frac{\sinh(\beta m/2 -
v_l/2)}{\sinh \beta m/2}\right) \nn\\
&\stackrel{\beta\rightarrow \infty}{\longrightarrow}&\ \
\exp\left(-\frac{1}{2}\,\sign(m)\sum_l v_l\right)\ee
%
where we assume that $\omega$ has compact support in taking the
limit in the last line. Translating back to Minkowski space, we
can write this as a contribution to the effective Lagrangian
involving the original $u(k)$ connection $\omega$.
\be L_{\rm eff} = \frac{1}{2}\,\sign(m)\,(\Tr\,\omega_a) \dot{X}^a
\label{yes}\ee
This is the promised result. We see that, even in the limit
$m\rightarrow \infty$, the zero modes leave a remnant of their
existence by inducing an effective magnetic field ${\cal F}=d{\cal
A}$ on the moduli space, where ${\cal A} = \Tr\,{\omega}$ is
defined in terms of the connection on the index bundle. ${\cal F}$
is proportional to the first Chern character of the index bundle
while, conveniently enough, ${\cal A}$ is known as the
Chern-Simons 1-form. This is the worldline counterpart to the
statement that the parent three-dimensional fermions induce a
Chern-Simons term.

\para
The result \eqn{yes} holds for integrating out the zero modes
associated to a single chiral multiplet fermion. As we saw in
\eqn{nkappa}, we must integrate $\tilde{N}=2\kappa$ chiral
multiplets. Our final result is that the low-energy dynamics of
vortices in the Chern-Simons theory \eqn{lag} is given by,
\be L = \frac{1}{2}g_{ab}\dot{X}^a\dot{X}^b - \kappa {\cal A}_a
\dot{X}^a \label{final}\ee

\subsubsection*{Why Only Zero Modes Matter}

In deriving the Lagrangian \eqn{final}, we have integrated out
only the zero modes on the vortex worldline, while ignoring the
infinite tower of higher solutions to the Dirac equation. We now
show that this is consistent. The key point is that higher
excitations of fermions come in pairs, with energy $\pm E$:
\be \left(\begin{array}{cc} 0 & i{\cal D}_z \\ -i{\cal
D}_{\bar{z}} & 0 \end{array}\right)\left(\begin{array}{c}
\tilde{\psi}_- \\  \tilde{\psi}_+\end{array}\right) =  E
\left(\begin{array}{c} \tilde{\psi}_- \\
\tilde{\psi}_+\end{array}\right) \ \ \Rightarrow\ \ \
\left(\begin{array}{cc} 0 & i{\cal D}_z \\ -i{\cal D}_{\bar{z}} &
0 \end{array}\right)\left(\begin{array}{c} \tilde{\psi}_- \\
-\tilde{\psi}_+\end{array}\right) =  -E
\left(\begin{array}{c} \tilde{\psi}_- \\
-\tilde{\psi}_+\end{array}\right) \nn\ee
Contributions to the  Chern-Simons term on the vortex
worldline cancel between each pair. To see this, we write
the general eigenfunction as $\tilde{\psi}^T =
(\tilde{\psi}_-\zeta_-, \tilde{\psi}_+\zeta_+)$ and promote
$\zeta_\pm$ to time-dependent Grassmann fields. The action for
these objects is schematically
\be L_{\rm non-zero-modes} = \bar{\zeta}_+(iD_t-m)\zeta_+ +
{\bar{\zeta}}_-(iD_t+m){\zeta}_- + E({\bar{\zeta}}_+{\zeta}_- +
\bar{\zeta}_-\zeta_+)\ee
which is schematic only in the sense that we have dropped overall
coefficients that arise from the overlap of the eigenfunctions.
Integrating out the non-zero modes now gives us a determinant of
the form,
\be \det\left(\begin{array}{cc} iD_t -m & E \\ E &
iD_t+m\end{array}\right)= \det(iD_t +\sqrt{m^2+E^2})\det(iD_t -
\sqrt{m^2 + E^2})\ \ \ \ \ \ee
We see that the effective mass of these objects is
$\pm\sqrt{m^2+E^2}$, leading to a cancellation due to the presence
of the  $\sign(m)$ term in \eqn{yes}. In the limit $m\rightarrow
\infty$, these non-zero modes leave no trace of their existence on
the vortex dynamics.

\subsection{Abelian Vortices and the Tangent Bundle}
\label{abstring}

Our final answer \eqn{final} for the vortex dynamics  is pleasingly
 simple and geometrical. Yet it suffers from two drawbacks.
Firstly, we have no concrete expression for the index bundle and
its associated first Chern character. Secondly, as mentioned in
footnote \ref{footnote}, there is a technical subtlety due to the
non-normalizability of the zero modes. In this section we remedy
both of these issues for Abelian vortices. In Section \ref{doit}
 we shall also remedy the problem of non-normalizability for
non-Abelian vortices.

\para
Our strategy is a slightly more refined version of that described
above. We again generate the Chern-Simons terms by integrating out
supplementary matter multiplets. The only thing that differs from
the previous discussion is the matter that we choose to integrate
out. Our starting point this time will be the Abelian Higgs model
with ${\cal N}=4$ supersymmetry (i.e. 8 supercharges). We set
$\kappa =0$ in \eqn{lag}, and introduce a neutral chiral multiplet
$A$, containing the Dirac fermion $\eta$, together with a single
chiral multiplet $\tilde{Q}$ of charge $-1$, containing the
fermion $\tilde{\psi}$. The extended supersymmetry requires that
these are coupled to the original chiral multiplet $Q$, containing the
scalar $q$, through the superpotential,
\be {\cal W} = \sqrt{2}\tilde{Q}AQ\label{w}\ee
The benefit of working in the ${\cal N}=4$ model is that the
geometry of the fermi zero modes is well understood. Indeed, in
the background of the Abelian vortex, the Dirac equations for
$\eta$ and $\tilde{\psi}$ reduce to\footnote{A recent
detailed discussion of these issues, with an explicit
demonstration of the relationship between fermionic and bosonic
zero modes, can be found in \cite{het}.},
%
%
%
\be i{\cal D}_{\bar z} \eta_- -
\sqrt{2}q^\dagger\tilde{\psi}^\dagger_- = 0\ \ \ ,\ \ \ -i{\cal
D}_z\tilde{\psi}^\dagger_--\sqrt{2}\eta_-q=0
\label{paul}\ee
%
The index theorem remains the same as before, and
these equations again have $k$ complex zero modes. However,
the presence of the coupling to $q$
--- which has a non-zero vacuum expectation value --- ensures that
the zero mode profiles are localized exponentially near the vortex
cores and are normalizable. This resolves the problem described in
footnote \ref{footnote}.

\para
Moreover, it can be shown that the $k$
fermi zero modes are proportional to the bosonic zero modes of the
vortex: they are related by the extended supersymmetry.
%
%
The upshot of this is that the fermi zero modes live
--- like their bosonic counterparts --- in the {\it tangent bundle}
over ${\cal M}$. The appropriate
covariant derivative for the $k$ Grassmann collective coordinates
$\xi$ is now,
\be (D_t\xi)^a = \partial_t\xi^a + \Gamma^a_{bc}\dot{Z}^b\xi^c\ee
where, in contrast to previous formulae, we have switched to
complex notation, defining the holomorphic coordinates $Z^a$,
$a=1,\ldots, k$ on a patch of the moduli space ${\cal M}$. The
$\Gamma^a_{bc}$ are the holomorphic components of the Levi-Civita
connection.

\para
The above is merely a review of well known results about fermi
zero modes of vortices in theories with ${\cal N}=4$
supersymmetry. As before, we now deform our theory by adding
a real mass $m$ for the chiral
multiplets $A$ and $\tilde{Q}$. We then integrate $A$ and $\tilde{Q}$
out. The multiplet $A$ is neutral and decouples in the $m\rightarrow
\infty$ limit. In contrast, $\tilde{Q}$ induces a Chern-Simons
interaction with coefficient $\kappa = -\ft12\, \sign(m)$.

\para
Integrating out the fermi zero modes on the worldline proceeds as
before. But, since the zero modes live in the tangent bundle,
locally we have $d{\cal A} = R$ where $R$ is the Ricci form. This
is defined in terms of the metric $g_{a\bar{b}}$ by
\be R = i\partial\bar{\partial}\ln\sqrt{g}\ee
In terms of local complex coordinates on ${\cal M}$, the vortex
dynamics becomes
\be L = g_{a\bar{b}}\dot{Z}^a\dot{\bar{Z}}{}^{\bar{b}} -
{\kappa}\left( {\cal A}_a\dot{Z}^a + \bar{\cal
A}_{\bar{a}}\dot{\bar{Z}}{}^{\bar{a}}\right)\label{evenmorefinal}\ee
where the complex Chern-Simons 1-form can be written locally as
\be {\cal A}_a = -\frac{i}{2}\,\frac{\partial}{\partial Z^a}\ln
\sqrt{g}\label{A}\ee

\subsection{Non-Abelian Vortices Revisited}
\label{doit}

The discussion in Section \ref{abstring} was solely for Abelian vortices. What goes
wrong if we try to repeat it for non-Abelian vortices? In order to
build the non-Abelian theory with ${\cal N}=4$ supersymmetry, we
must augment the $\kappa=0$ Lagrangian with $N$ chiral
multiplets $\tilde{Q}$ in the anti-fundamental representation, and
a single chiral multiplet $A$ in the adjoint representation.
Integrating out the $\tilde{Q}$ results in a $U(N)$ Chern-Simons
interaction of the type given in \eqn{lag}. However, integrating
out the adjoint multiplet $A$ contributes to the $SU(N)$
Chern-Simons term, but not the $U(1)$ Chern-Simons term. Thus the
mass deformed ${\cal N}=4$ theory does not yield the $U(N)$ ${\cal
N}=2$ theory of the form \eqn{lag}, but rather a theory with
different Chern-Simons coefficients for the $SU(N)$ and $U(1)$
parts of the gauge group.

\para
To make progress, we could instead augment the $\kappa=0$ Lagrangian
with $N$ chiral multiplets $\tilde{Q}$ in
the anti-fundamental representation, and a single neutral chiral
multiplet $A$. The theory no longer admits ${\cal N}=4$
supersymmetry, so we cannot use the above argument to show that
the zero modes live in the tangent bundle. Nonetheless, adding a
superpotential of the form \eqn{w} means that the Dirac equations
are once more of the form \eqn{paul}, and the fermi zero modes are
rendered normalizable. Thus, although we cannot show that the
magnetic field on the moduli space of non-Abelian vortices takes
the simple form \eqn{A}, any lingering worries caused by footnote
\ref{footnote} may now be left behind.

\section{Examples}

In this section, we illustrate our result with two examples. We
firstly examine the qualitative dynamics of two Abelian vortices
and show that the moduli space dynamics correctly captures their
fractional statistics. Secondly, we look at a single vortex in the
$U(N)$ theory, for which the internal moduli space is ${\bf
CP}^{N-1}$. We derive the dynamics both from the method described
in Section 3, and also from a direct moduli space computation.

\subsection{Two Abelian Vortices}

The relative dynamics of two Abelian vortices takes place in the
moduli space ${\cal M}\cong {\bf C}/{\bf Z}_2$. The metric is
given by
\be ds^2 = f^2(\sigma)(d\sigma^2 + \sigma^2d \theta^2)\ee
\EPSFIGURE{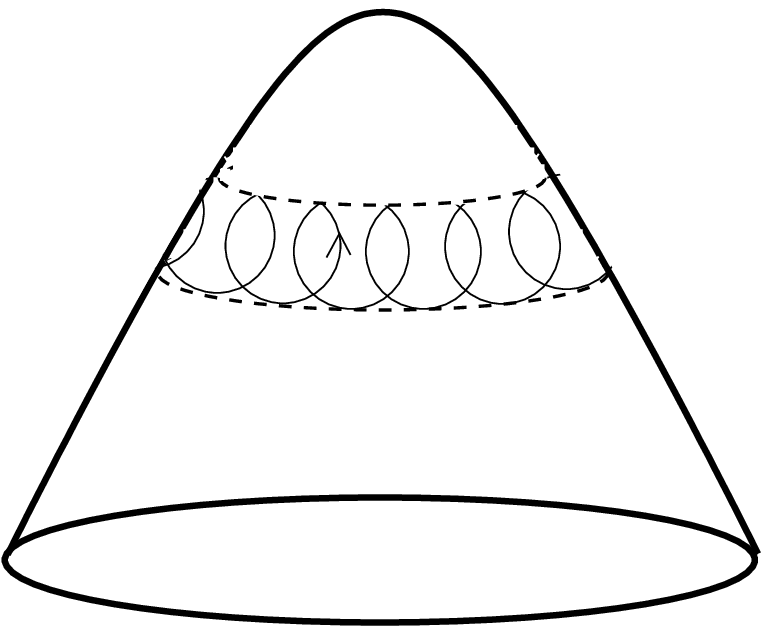,width=120pt}{The moduli space is a cone.}
\noindent where $\theta \in [0,\pi)$. Asymptotically, as $\sigma
\rightarrow \infty$, we have $f^2(\sigma)\rightarrow 1 +{\cal
O}(e^{-\sigma})$ \cite{samols,mansp} and the moduli space is a
cone with deficit angle $\pi$. Although the function $f(\sigma)$
is not known analytically, it can be shown that $f^2(\sigma) \sim
\sigma^2$ as $\sigma\rightarrow 0$, ensuring that the tip of the
cone is smooth. The moduli space is sketched in Figure 1, together
with an example of the motion which we will describe shortly.

\para
We work with the single valued holomorphic coordinate
$z=\sigma^2e^{2i\theta}$. Then the Chern-Simons 1-form \eqn{A} on the vortex
worldline is given by,
\be L_{CS} = -\kappa ({\cal A}\dot{z} + \bar{\cal A}\dot{\bar{z}})
= -\kappa \left(\frac{\sigma}{2}
\frac{\partial}{\partial\sigma}\log\,f^2 -
1\right)\dot{\theta}\label{twov}\ee
A similar expression, expressed in slightly different variables,
can be found in equation (85) of \cite{kimyeong}.

\para
Although the explicit function $f(\sigma)$ is not known, we may
still study the qualitative behaviour of vortices. The conserved
Noether charge associated to $\theta$ is given by
\be J = f^2\sigma^2\dot{\theta}
+\kappa\left(1-\frac{\sigma}{2}\frac{\partial\log
f^2}{\partial\sigma}\right)\label{j}\ee
As explained in \cite{kimyeong}, this differs from the angular
momentum of the two vortices by a constant. Meanwhile the
conserved Hamiltonian is
\be H = \frac{1}{2}f^2\dot{\sigma}^2 + V_{\rm eff}(\sigma)\ee
where the effective potential is due to the Chern-Simons term,
together with the usual angular momentum barrier,
\be V_{\rm eff}(\sigma) =
\frac{1}{2f^2\sigma^2}\left(J-\kappa+\frac{\kappa\sigma}{2}
\frac{\partial\log f^2}{\partial\sigma}\right)^2\ee
The classical scattering of vortices depends on the form of
$V_{\rm eff}$ which, in turn, depends on the relative values of
$\kappa$ and $J$. Let us fix $\kappa
>0$. On physical grounds, the form of the effective potential is shown in Figure 2:
\EPSFIGURE{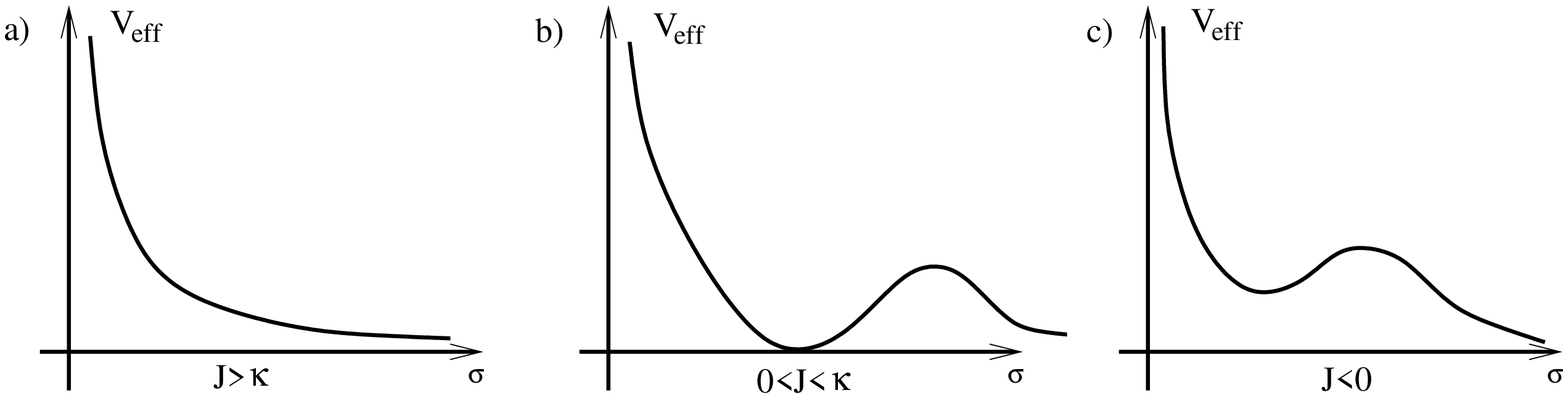,width=380pt}{The effective potential for different values of $J$.}
\begin{itemize}
\item $J>\kappa$. In this regime, we have $\dot{\theta}>0$ and
$V_{\rm eff}$ is shown in Figure 2a. $V_{\rm eff}$ acts as an effective
angular momentum barrier and the scattering of vortices is not
qualitatively different from the case without a Chern-Simons term.
\item The regime $0 < J < \kappa$ is more interesting. The
effective potential is shown in Figure 2b.  The root of the
effective potential corresponds to the static solution. We see
that, as emphasized in \cite{kimyeong}, static solutions with
different vortex separation $\sigma$ carry different angular
momentum $J$.

Small oscillations around the  minimum of $V_{\rm eff}$ give rise
to bound orbits of vortices. From the expression \eqn{j}, we see
that $\dot{\theta}$ oscillates from negative to positive in such
orbits. The corresponding motion on the moduli space is drawn in
Figure 1. The two vortices trace Larmor circles, while orbiting one another.
This moduli space motion can be understood using a standard
argument involving adiabatic invariants: in the slowly varying
magnetic field, a particle drifts along
lines of constant field strength.
\item For $J<0$, we have $\dot{\theta}<0$. There are two distinct
shapes of $V_{\rm eff}$. For suitably small $|J|$, the effective
potential takes the form shown in Figure 2c. There are once again
bound orbits, including one at fixed $\sigma$. For $J\ll 0$, the
minimum of $V_{\rm eff}$ disappears and the potential once again
takes the shape of Figure 2a, with only scattering trajectories.
\end{itemize}
Before we move on, we also note that there is a simple quantum
effect that follows from \eqn{twov}. The first term vanishes as
$\sigma\rightarrow \infty$, while the second survives.  This
ensures that as the particles orbit asymptotically, the
wavefunction picks up a phase $\exp(\pm i\pi\kappa)$. For $\kappa
\nin{\bf Z}$, this endows the vortices with fractional statistics
in agreement with the analysis of
\cite{csdyn,kimyeongold,kimyeong}.

\subsection{One Non-Abelian Vortex}

For our second example, we examine a single vortex in $U(N)$. We
first review the dynamics of the vortex in the $\kappa =0$ case.
The vortex has an internal moduli space ${\cal M}\cong{\bf
CP}^{N-1}$, describing its orientation in colour and flavour
space \cite{vib,auzzi}. We introduce homogeneous coordinates on ${\cal M}$ by
starting with a solution $B_\star$ for the magnetic field of a
single Abelian vortex configuration. We can embed the Abelian
solution into a non-Abelian configuration by writing,
\be B^a_{\ b} =
\frac{B_\star}{r}\,\varphi^a\bar{\varphi}_b\label{namby}\ee
with a similar expression for the Higgs field which we will
describe in more detail in Section \ref{itsthis}. The coordinates
$\varphi_a \in {\bf C}$, $a=1,\ldots, N$, satisfy the constraint,
\be \sum_{a=1}^N |\varphi_a|^2 = r \label{constraint}\ee
where $r$ is a constant which is determined to be $r = 2\pi/e^2$
\cite{vib,auzzi,gsy}.  The solutions \eqn{namby} are invariant
under the simultaneous rotation
\be \varphi_a\rightarrow e^{i\vartheta}\varphi_a\label{iden}\ee
The $\varphi_a$, subject to the constraint \eqn{constraint} and
identification \eqn{iden}, provide homogeneous coordinates on the
moduli space ${\cal M}\cong{\bf CP}^{N-1}$. The low-energy
dynamics of the vortex is described by a sigma-model on ${\cal M}$
endowed with the Fubini-Study metric and K\"ahler class $r$.
There is a simple way to impose the identification \eqn{iden} by
introducing an auxiliary gauge field $\alpha$ on the worldline.
The Lagrangian for the internal modes of the vortex takes the
form,
\be L_{vortex} = \sum_{a=1}^N |{\cal D}_t\varphi_a|^2\ee
where the degrees of freedom are subject to the constraint
\eqn{constraint}, and the covariant derivative is given by ${\cal
D}_t\varphi_a = \dot{\varphi}_a - i\alpha\varphi_a$.

\para
Let us now ask how this dynamics is altered by the presence of the
Chern-Simons term. The moduli space is compact and
the cohomology is generated by $\Omega$, the K\"ahler form. Thus
the first Chern character ${\cal F}$ of the
index bundle must be proportional to $\Omega$. We need only
determine the proportionality constant. In fact, this is simple to
achieve in the language introduced above. Let $\tilde{\psi}_\star$
denote the solution to the Abelian Dirac equation \eqn{dirac}.
Then the solution to the non-Abelian Dirac equation, with gauge
field given by \eqn{namby}, is
\be \tilde{\psi}^b  =  \tilde{\psi}_\star\ \xi\bar{\varphi}_b \ee
This is compatible with the symmetry \eqn{iden} if the Grassmann
collective coordinate $\xi$ is assigned charge,
\be \xi \rightarrow e^{i\vartheta}\xi\ee
This transformation rule determines the index
bundle, for it fixes the kinetic term of the Grassmann variable to
be given by the covariant derivative $D_t\xi =
\dot{\xi}-i\alpha\xi$. We may now take $m\rightarrow \infty$, and
integrate out $\xi$. The calculation is the same as that
described in Section 3, and yields
\be L_{1-vortex} = \sum_{a=1}^N|{\cal D}_t\varphi_a|^2 - \kappa
\alpha\ee
%

%
%

\subsubsection*{An Example of the Example}

For a single vortex in the $U(2)$ theory, the moduli space is
${\bf S}^2\cong{\bf CP}^1$. We now provide a more explicit
description of the dynamics in this case. The constraints
\eqn{constraint} are simply solved by
\be \varphi_1 = \sqrt{r} e^{i\psi-i\phi/2}\cos(\theta/2) \ \ \ ,\
\ \ \varphi_2 = \sqrt{r}
e^{i\psi+i\phi/2}\sin(\theta/2)\label{coords}\ee
where the angles take ranges $\psi\in [0,2\pi)$, $\phi\in
[0,2\pi)$ and $\theta\in[0,\pi)$. Expanding out the Lagrangian
gives
\be L_{vortex} =
r\cos^2(\theta/2)(\dot{\psi}-\dot{\phi}/2-\alpha)^2 + r
\sin^2(\theta/2)(\dot{\psi}+\dot{\phi}/2 - \alpha)^2 +
\frac{r}{4}\dot{\theta}{}^2 - \kappa\alpha\nn\ee
We now eliminate the gauge field $\alpha$ by its equation of
motion. Ignoring an overall constant term and treating total derivatives carefully, the resulting dynamics
is given by
\be L_{1-vortex} = \frac{r}{4}\left[\dot{\theta}{}^2+
\sin^2\theta\,\dot{\phi}{}^2\right] +
\frac{\kappa}{2}(\cos\theta-1)\dot{\phi}\label{mono}\ee
We recognize the first term as the familiar sigma-model on ${\bf S}^2$
with radius $R = \sqrt{r/2}$. The second term is the Dirac
monopole connection of strength $\kappa$, expressed in a form which
gives a well-defined potential everywhere except at the south pole.

\subsection{One Non-Abelian Vortex: Explicit Moduli Space Computation}
\label{itsthis}

In this final section, we show how to re-derive the Dirac monopole
connection \eqn{mono} from an explicit moduli space calculation. As
we shall see, the calculation requires that we take care with the
topology of the moduli space.

\para
Following \cite{kimyeong}, we work perturbatively both in the
velocity of the vortices, and in $\kappa$. Practically, this means
that we start with the Bogomolnyi equations with
$\kappa=0$,
\be B=\frac{e^2}{2}(q_iq_i^\dagger - v^2)\ \ \ ,\ \ \ {\cal D}_z q_i=0
\label{vortex}\ee
but with $\phi=A_0$ determined by Gauss' law
\eqn{gauss}\footnote{Since $\kappa \in {\bf Z}$, it does not seem
like a good candidate for perturbation theory. A more careful
study shows that $e^2\kappa^2/v^2 \ll 1$ is the small parameter.}.
Let's first quantify the price that we pay by working
perturbatively in $\kappa$. Since the Chern-Simons term clearly
plays a crucial role in this discussion, it is necessary to  work
with the Lagrangian instead of the energy functional. We evaluate
the Lagrangian \eqn{lag} on the solution to the equations
\eqn{vortex} and \eqn{gauss}, with $\partial_0=0$. This gives
%
%
%
\be L = \int d^2x {\cal L} = -2\pi v^2 k -
\frac{e^2\kappa^2}{16\pi^2} \int\ d^2x \Tr\,\phi^2\ee
The last term is the correction to the Lagrangian due to the fact
that we chose to work with the $\kappa=0$ Bogomolnyi equations,
rather than the true equations \eqn{bog1} and \eqn{bog2}.
The mass of the configuration is
\be M_{\rm vortex} = 2\pi v^2 k \left( 1+ {\cal
O}\left(\frac{e^4\kappa^4}{v^4}\right)\right)\ee
The extra term is the price we pay for our approximation. At our
level of approximation, we  neglect all terms of this order in
what follows.

\subsubsection*{Zero Modes}

Let us now turn to the dynamics of the system. Here we see the advantage
of our approximation, because we may deal with
the familiar vortex equations \eqn{vortex}. Denote the collective
coordinates of this system by $X^a$, with $a=1,\ldots 2kN$. The
zero modes of the solution are then given by differentiating, together
with a gauge transformation:
\be \delta_aA_\alpha = \frac{\partial A_\alpha}{\partial
X^a}-{\cal D}_\alpha w_a\ \ \ ,\ \ \ \delta_a q_i =
\frac{\partial{q_i}}{\partial X^a} - iw_aq_i\label{zeromodes1}\ee
The gauge transformation $w_a\in u(N)$ is dictated by the gauge fixing
condition,
\be {\cal D}_\alpha \delta_a A_\alpha =
-\frac{ie^2}{2}(\delta_aq_i\,q_i^\dagger -
q_i\delta_aq_i^\dagger)\label{cons}\ee
We next write $A_0=w+\phi$, where $w \equiv w_a\dot{X}^a$,
which ensures that the zero modes are related to the covariant time
derivatives as follows:
\be {\cal D}_0q_i = \delta_aq_i\dot{X}^a -i\phi q_i\  \ \ \ \ ,\ \
\ E_\alpha = \delta_aA_\alpha \dot{X}^a - {\cal D}_\alpha\phi\label{D5-4.20}\ee
The presence of the $\phi$ terms on the right-hand-side of these
equations is what distinguishes the Chern-Simons dynamics from the
case $\kappa=0$. Notice that in our approximation, we have not
needed to linearize the second order Gauss' law equation
\eqn{gauss} since the terms $({\cal D}_0\phi)^2$ are of order
$\kappa^2\dot{X}^2$ and may be safely ignored. Substituting into											 %
the Lagrangian \eqn{lag}, and making use of the constraint \eqn{cons}, we
derive an expression for the Lagrangian governing the dynamics of
the vortex,
\be L =
g_{ab}\dot{X}^a\dot{X}^b - 2\pi v^2 k
- \frac{\kappa}{4\pi}\int d^2x\ \Tr\, \left(2B w_a\dot{X}^a
-\epsilon_{\alpha\beta}A_\alpha\dot{A}_\beta\right) \label{kv}\ee
This generalizes the result derived in
\cite{csdyn,kimyeongold,kimyeong} to the non-Abelian case. The
first term in this expression is the usual metric on the vortex
moduli space, given by
\be g_{ab} = \int d^2x\ \left( \frac{1}{e^2} \Tr\,
\delta_aA_\alpha\delta_b A_\alpha +
\delta_{(a}q_i{}^{\!\dagger} \delta_{b)}q_i\right)\ee
The effect of the Chern-Simons interaction is shown in the last term
of \eqn{kv}, which is of order $\kappa\dot{X}$.

\subsubsection*{Non-Singular Gauge}

We now apply this formula to the simple case of a single vortex in
the $U(2)$ gauge theory. In this case, the moduli space is ${\bf
CP}^1$. Previous field theoretic studies of this system have
always employed singular gauge \cite{auzzi}, in which the Higgs
field $q_i$ has no winding at infinity. While this gauge is
perfectly adequate for studying the metric on moduli space (see,
for example \cite{gsy}), it hides the interesting topology of the
moduli space and is not suitable for studying the effect of the
Chern-Simons term.
We therefore first describe the collective coordinates of the
single $U(2)$ vortex in a gauge that does not suffer from singular
behaviour.

\para
Consider the $U(1)$ vortex equations \eqn{vortex}. We work in
polar coordinates on the spatial plane: $x_1= \rho \cos\chi$ and
$x_2=\rho\sin \chi$. Then the solution to the equations for the
$k=1$ vortex is given by
\be q = vq_\star(\rho) e^{i\chi}\ \ \ {\rm and}\ \ \ A_\chi =
1-f(\rho)\ \ \ ,\ \ \ A_\rho = 0 \ee
where the profile functions satisfy the ordinary differential
equations,
\be \rho q_\star^\prime = -fq_\star\ \ \ {\rm and}\ \ \
\frac{f^\prime}{\rho}=-\frac{e^2v^2}{2}(q_\star^2-1)\ee
subject to the boundary conditions $q_\star(\rho)\rightarrow 1,0$ and
$f(\rho)\rightarrow 0,1$ as $\rho \rightarrow +\infty, 0$. Given
these Abelian solutions, it is now a simple matter to embed them
into the fields of the $U(2)$ theory to arrive at new solutions.
There are two natural embeddings:
\be &i)&\ \ \ \ q_{(1)} = v\left(\begin{array}{cc} q_\star
e^{i\chi} & 0 \\ 0 & 1 \end{array}\right) \ \ \ ,\ \ \ A_\chi =
\left(\begin{array}{cc} (1-f) & \ 0 \\ 0 &\  0
\end{array}\right) \ \ \ ,\ \ \  A_\rho = 0\label{q1}\\ &ii)&\ \ \ \
q_{(2)} = v\left(\begin{array}{cc} 1 & 0 \\ 0 & q_\star e^{i\chi}
\end{array}\right) \ \ \ ,\ \ \ A_\chi = \left(\begin{array}{cc}
0\   & 0 \\ 0\  & (1-f) \end{array}\right) \ \ \ ,\ \ \  A_\rho =
0 \label{q2}\ee
Here the rows and columns of the $q$ matrix correspond to colour
and flavour indices respectively. However, these embeddings are
not the only two. Given either of these solutions, one may act
upon it with a diagonal combination of the $SU(2)_{\rm flavour}$
symmetry and $SU(2)_{\rm gauge}$ symmetry of the model in such a
way that the diagonal structure of the vacuum remains invariant,
\be q \rightarrow U qV^\dagger\ \ \ ,\ \ \ A\rightarrow
UAU^\dagger - i (\partial U)\,U^\dagger\ee
where $V\in SU(2)_{\rm flavour}$ is a constant matrix, and
$U=U(\rho,\chi)\in SU(2)_{\rm gauge}$. In singular gauge, we would
impose the condition that $U\rightarrow V$ as $\rho\rightarrow
\infty$. However, the presence of the winding scalar field in
\eqn{q1} and \eqn{q2} means that cannot be quite right in the
present case. Indeed, the only transformation such that
$U\rightarrow V$ that is allowed is
$U=V=\tiny{\left(\begin{array}{cc} 0 & i
\\ i & 0
\end{array}\right)}$ which maps $q_{(1)}$ to $q_{(2)}$. For more
general transformations, $U$ must itself include some winding. The
necessary condition is not difficult to determine. For
$V=\tiny{\left(\begin{array}{cc} \hat{a}_1 & \hat{a}_2 \\
\hat{a}_3 & \hat{a}_4\end{array}\right)} \in SU(2)_{\rm flavour}$,
we require
\be U_{(1)}(\rho,\chi) = \left(\begin{array}{cc} a_1(\rho) &
a_2(\rho)e^{i\chi} \\ a_3(\rho)e^{-i\chi} &
a_4(\rho)\end{array}\right)\ \ \ {\rm or}\ \ \ U_{(2)}(\rho,\chi)
= \left(\begin{array}{cc} a_1(\rho) & a_2(\rho)e^{-i\chi} \\
a_3(\rho)e^{i\chi} & a_4(\rho)\end{array}\right) \ \ \ \ \ \ \
\label{u12}\ee
where the matrix $U_{(1)}$ is to be used for transformations away
from $q_{(1)}$, while the matrix $U_{(2)}$ is required for
transformations away from $q_{(2)}$. In both cases, the profile
functions in the gauge transformation satisfy the boundary
conditions $a_i(\rho)\rightarrow \hat{a}_i$ as $\rho \rightarrow
\infty$.

\para
Perhaps unsurprisingly, the picture that emerges is that two
patches are required to cover the moduli space. The solution
$q_{(1)}$ can be thought of as the north pole of ${\bf CP}^1$, and
combined gauge and flavour transformations given by $U_{(1)}$
cover nearly all the space, but cannot take us to $q_{(2)}$.
Similarly, $q_{(2)}$ is thought of as the south pole of the moduli
space and transformations using $U_{(2)}$ can reach the full
moduli space, except for the north pole.

\subsubsection*{Finding the Dirac Monopole Connection}

We now use these results to derive
the Dirac monopole connection on moduli space. Let's start with
the solution $q=q_{(1)}$. We look for zero modes corresponding to
a simultaneous $SU(2)$ gauge and flavour rotation, with parameters $\Omega$ and $\hat{\Omega}$ respectively. The zero modes are given
by
\be \delta q \equiv \delta_a q \dot{X}^a = i(\Omega q - q\hat{\Omega})\ \ \ , \ \ \ \delta A_\alpha \equiv \delta_a A_\alpha \dot{X}^a= {\cal D}_\alpha \Omega \label{zmo}\ee
The requirement that the vacuum remains invariant fixes $\Omega_\infty \equiv \lim_{\rho \rightarrow \infty} \Omega(\rho,\chi)$ in terms of $\hat{\Omega}$.  The remaining freedom in $\Omega$ is fixed by the constraint \eqn{cons}, which now reads
\be {\cal D}^2\Omega =
\frac{e^2}{2}\left(\{\Omega,qq^\dagger\}-2q\hat{\Omega}q^\dagger\right)\label{omega}\ee
We demand that varying the fields with respect to the collective co-ordinates corresponds to the `large' part of the gauge and flavour rotation, with parameters $\Omega_\infty$ and $\hat{\Omega}$.  This means that
\be  \partial_0 q = \frac{\partial q}{\partial X^a} \dot{X}^a = i\left( \Omega_\infty q - q \hat{\Omega} \right)\ \ \ {\rm and} \ \ \ \partial_0 A_\alpha = \frac{\partial A_\alpha}{\partial X^a} \dot{X}^a = {\cal D}_\alpha \Omega_\infty \label{D6-4.31}\ee
To achieve this and satisfy \eqn{zmo}, we set $w = \Omega_\infty - \Omega$ in \eqn{zeromodes1}.
\para
%

We choose our flavour transformation to be $\hat{\Omega} =
\frac{\dot{\theta}}{2}\tiny{\left(\begin{array}{cc} 0 & 1 \\ 1 & 0
\end{array}\right)}\in su(2)_{\rm flavour}$, where we are using
the coordinates \eqn{coords}, and the factor of $\theta/2$ in this
expression follows directly from the same factor in \eqn{coords}.
Then \eqn{omega} is solved by \cite{gsy}
\be \Omega = \frac{\dot{\theta}}{2}\left(\begin{array}{cc} 0 & q_\star(\rho) e^{i\chi} \\
q_\star(\rho) e^{-i\chi} & 0 \end{array}\right)\label{431}\ee
where the boundary condition on $\Omega$ is inherited from
$\hat{\Omega}$. The asymptotic winding in \eqn{431} results from
working in non-singular gauge as in equation \eqn{u12}.
\para
To compute the terms in the low-energy dynamics of the Lagrangian, we substitute \eqn{D6-4.31} and $w = \Omega_\infty - \Omega$ into our moduli dynamics \eqn{kv} to get
\be L_{\rm CS} = -\frac{\kappa}{4\pi} \int d^2x \ \Tr\,\left( 2B(\Omega_\infty - \Omega) - \epsilon_{\alpha\beta}A_\alpha {\cal D}_\beta \Omega_\infty \right) \label{D6-4.33}\ee
Using \eqn{431} and \eqn{q1}, we see that $(\Omega_\infty - \Omega)$ and ${\cal D}_\rho \Omega_\infty$ are off-diagonal, while $B$ and $A_\chi$ are diagonal and $A_\rho$ is zero.  Hence \eqn{D6-4.33} vanishes.

\para
\EPSFIGURE{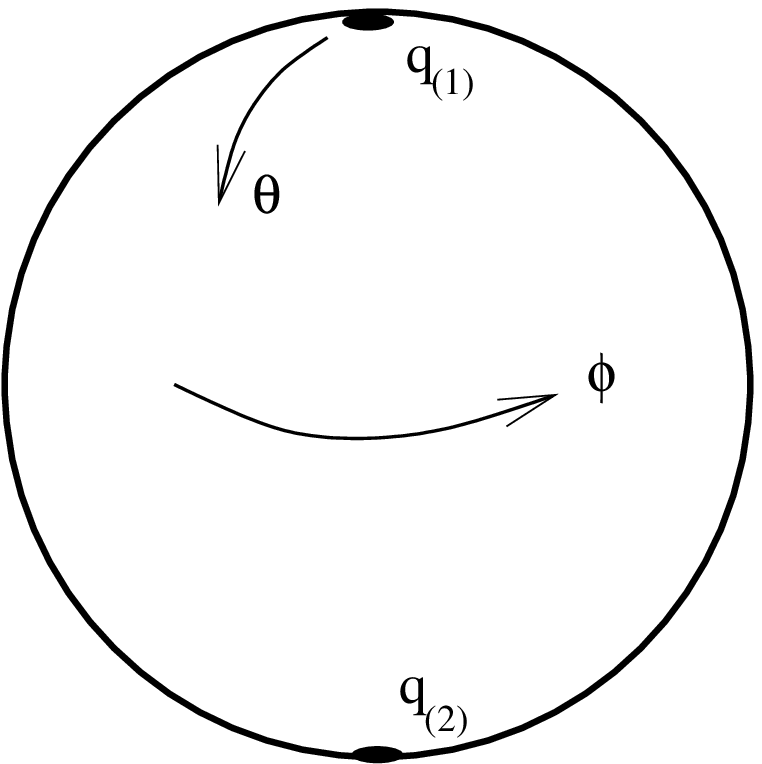,width=100pt}{\ }
However, we shouldn't be too hasty in concluding that the
Chern-Simons term has no effect on the vortex dynamics.  We should
first compare with the expected Dirac monopole solution found in
Section 4.2. We have worked about the ``north pole" solution
\eqn{q1}. As discussed previously, this patch covers all but the
south pole of ${\bf CP}^1$. If we were to write the Dirac monopole
in these coordinates, the Dirac string would point along the
direction of the south pole. The corresponding term on the
worldline is given by
\be L_{\rm Dirac} = \frac{\kappa}{2}(\cos\theta-1)\dot{\phi}\ee
However, as shown in Figure 3, the calculation that we have just
done corresponds to moving downwards from the north pole. This is
equivalent to looking for a $\dot{\theta}$ term in the effective
action. It is not surprising that it gave a vanishing answer! Said
another way, there is always a coordinate choice so that a given
infinitesimal motion doesn't reveal a Dirac monopole connection in
the Lagrangian. We have made that coordinate choice above; moreover,
such a coordinate choice is always made implicitly if we work in singular gauge because this
gauge disguises the presence of the Dirac string.							

\para
With this understanding of the topology of moduli space, it is a
simple matter to perform a calculation that does see the Dirac
monopole connection. Our first goal is to rotate the $q_{(1)}$
solution to a configuration corresponding to latitude $\theta$ on
the moduli space. This is done by a flavour rotation of the form,
\be V = \left(\begin{array}{cc} \cos(\theta/2) & i\sin(\theta/2)
\\ i\sin(\theta/2) & \cos(\theta/2)\end{array}\right) \in
SU(2)_{\rm flavour}\ee
together with a suitable gauge transformation $U_{(1)}$ with
boundary conditions given in \eqn{u12}. We now search for zero
modes around this new background. Our task is to solve for the
infinitesimal gauge transformation $\Omega$ satisfying
\eqn{omega}, subject to the appropriate
boundary condition. This boundary condition comes
from the requirement that the gauge transformation acts in the
longitudinal $\phi$ direction, and returns us to our starting point after $\phi$ has increased by $2\pi$. Using the coordinates
\eqn{coords}, we see that this can be achieved if we supplement our gauge and flavour transformations by $U(1)$ rotations corresponding to motion in the $\psi$ direction.  An appropriate choice is
$\hat{\Omega} = \dot{\phi}\tiny{\left(\begin{array}{cc} 0 & 0 \\
0 & 1
\end{array}\right)}$.  Since this is diagonal, we have $\Omega_\infty = \hat{\Omega}$ (see \eqn{u12}).  Using the fact that $\partial_\alpha \Omega_\infty = 0$ and performing an integration by parts, we may write
\be \int d^2x\ \Tr\,\left(-\epsilon_{\alpha\beta}A_\alpha {\cal D}_\beta \Omega_\infty \right) = \int d^2x\ \Tr\,\left(-2B\Omega_\infty \right) \ee
Once we substitute this into the Lagrangian \eqn{D6-4.33}, we are left with
\be L_{\rm CS} = -\frac{\kappa}{4\pi} \int d^2x \ \Tr\,\left( -2B\Omega \right) \label{D6-LCS}\ee
%


\para
We may make use of the gauge covariance of \eqn{omega} to
translate the task of finding $\Omega$ into something equivalent: solving
\eqn{omega} in the background of the original vortex solution
\eqn{q1} now subject to the boundary condition arising from
\be V^\dagger \hat{\Omega} V= \dot{\phi}\,V^\dagger
\left(\begin{array}{cc} 0 & 0 \\ 0 & 1 \end{array}\right)V =
\frac{\dot{\phi}}{2} \left(\begin{array}{cc} 1-\cos\theta & -i\sin\theta \\
i\sin\theta & 1+\cos\theta\end{array}\right)\ee
It is straightforward to show that the solution is given by
\be U^\dagger \Omega(\rho,\chi) U = \frac{\dot{\phi}}{2}
\left(\begin{array}{cc} 1-\cos\theta & -ie^{i\chi}q_\star(\rho)
\sin\theta \\ ie^{-i\chi} q_\star(\rho) \sin\theta &
1+\cos\theta\end{array}\right)\ee
%
%
We now substitute our results into the expression \eqn{D6-LCS}
arising from moduli space dynamics.  Noting that the magnetic field associated with \eqn{q1} is given by $U^\dagger BU$, we have
%
\begin{eqnarray*} L_{\rm CS} &= &\frac{\kappa}{2\pi}\int d^2x\ \Tr\,U^\dagger\Omega U U^\dagger B U \\
 &=&\frac{\kappa(1-\cos\theta) \dot{\phi}}{4\pi}  \int d^2x\ \Tr\, B =
\frac{\kappa}{2} (\cos\theta-1) \dot{\phi} \end{eqnarray*}
%
%
%
This reproduces the Dirac monopole connection as claimed.


\setcounter{section}{0} \setcounter{equation}{0}
\renewcommand{\thesection}{\Alph{section}}

\section*{Acknowledgement}
We would like to thank Nick Dorey, Maciej Dunajski and Nick Manton
for many useful discussions. BC is supported by an STFC studentship.
DT is supported by the Royal Society.


\begin{thebibliography}{99}

\small
\parskip=0pt plus 2pt

\bibitem{manton} N.~S.~Manton,
  ``{\it A Remark On The Scattering Of BPS Monopoles},''
  Phys.\ Lett.\  B {\bf 110}, 54 (1982).

\bibitem{cs}   S.~Deser, R.~Jackiw and S.~Templeton,
  ``{\it Topologically massive gauge theories},''
  Annals Phys.\  {\bf 140}, 372 (1982)
  [Erratum-ibid.\  {\bf 185}, 406.1988\ APNYA,281,409 (1988\ APNYA,281,409-449.2000)].

\bibitem{erick} E.~J.~Weinberg,
``{\em Multivortex Solutions Of The Ginzburg-Landau Equations}'',
Phys.\ Rev.\ D {\bf 19}, 3008 (1979);
 ``{\em Index Calculations For The Fermion - Vortex System}''
Phys.\ Rev.\ D {\bf 24}, 2669 (1981).

\bibitem{taubes} A.~Jaffe and C.~Taubes,
``{\em Vortices And Monopoles. Structure Of Static Gauge
Theories}'', Birkhaeuser (1980).

\bibitem{samols}   T.~M.~Samols,
  ``{\it Vortex Scattering},''
  Commun.\ Math.\ Phys.\  {\bf 145}, 149 (1992).

\bibitem{manchen}   H.~Y.~Chen and N.~S.~Manton,
  ``{\it The K\"ahler potential of Abelian Higgs vortices},''
  J.\ Math.\ Phys.\  {\bf 46}, 052305 (2005)
  [arXiv:hep-th/0407011].


\bibitem{mansp}   N.~S.~Manton and J.~M.~Speight,
  ``{\it Asymptotic interactions of critically coupled vortices},''
  Commun.\ Math.\ Phys.\  {\bf 236}, 535 (2003)
  [arXiv:hep-th/0205307].

\bibitem{csdyn}  S.~K.~Kim and H.~S.~Min,
  ``{\it Statistical Interactions Between Chern-Simons Vortices},''
  Phys.\ Lett.\  B {\bf 281}, 81 (1992).

\bibitem{kimyeongold} Y.~Kim and K.~M.~Lee,
  ``{\it Vortex dynamics in selfdual Chern-Simons Higgs systems},''
  Phys.\ Rev.\  D {\bf 49}, 2041 (1994)
  [arXiv:hep-th/9211035].

\bibitem{kimyeong} Y.~Kim and K.~M.~Lee,
  ``{\it First and second order vortex dynamics},''
  Phys.\ Rev.\  D {\bf 66}, 045016 (2002)
  [arXiv:hep-th/0204111].

\bibitem{vib}  A.~Hanany and D.~Tong,
  ``{\it Vortices, instantons and branes},''
  JHEP {\bf 0307}, 037 (2003)
  [arXiv:hep-th/0306150].

\bibitem{auzzi}   R.~Auzzi, S.~Bolognesi, J.~Evslin, K.~Konishi and A.~Yung,
  ``{\it Nonabelian superconductors: Vortices and confinement in N = 2 SQCD},''
  Nucl.\ Phys.\  B {\bf 673}, 187 (2003)
  [arXiv:hep-th/0307287].

\bibitem{gw}  J.~Goldstone and F.~Wilczek,
  ``{\it Fractional Quantum Numbers On Solitons},''
  Phys.\ Rev.\ Lett.\  {\bf 47}, 986 (1981).

\bibitem{cw}  Y.~H.~Chen and F.~Wilczek, ``{\it Induced Quantum
Numbers in Some (2+1)-Dimensional Models}", in  ``Fractional
statistics and anyon superconductivity,'', World Scientific (1990)

\bibitem{redlich} A.~N.~Redlich, ``{\it Gauge Noninvariance and
Parity Nonconservation of Three Dimensional Fermions}",
  Phys.\ Rev.\ Lett.\  {\bf 52}, 18 (1984);
  ``{\it Parity Violation And Gauge Noninvariance Of The Effective Gauge Field
  Action In Three-Dimensions},''
  Phys.\ Rev.\  D {\bf 29}, 2366 (1984).

\bibitem{agw} L.~Alvarez-Gaume and E.~Witten,
  ``{\it Gravitational Anomalies},''
  Nucl.\ Phys.\  B {\bf 234}, 269 (1984).

\bibitem{5dcs}   B.~Collie and D.~Tong,
 ``{\it Instantons, Fermions and Chern-Simons Terms},''
  arXiv:0804.1772 [hep-th].

\bibitem{llm}   C.~K.~Lee, K.~M.~Lee and H.~Min,
  ``{\it Selfdual Maxwell Chern-Simons solitons},''
  Phys.\ Lett.\  B {\bf 252}, 79 (1990).

\bibitem{hkp}  J.~Hong, Y.~Kim and P.~Y.~Pac,
  ``{\it On The Multivortex Solutions Of The Abelian Chern-Simons-Higgs
  Theory},'' Phys.\ Rev.\ Lett.\  {\bf 64}, 2230 (1990).


\bibitem{jw}  R.~Jackiw and E.~J.~Weinberg,
  ``{\it Selfdual Chern-Simons Vortices},''
  Phys.\ Rev.\ Lett.\  {\bf 64}, 2234 (1990).


\bibitem{jlw}   R.~Jackiw, K.~M.~Lee and E.~J.~Weinberg,
  ``{\it Selfdual Chern-Simons solitons},''
  Phys.\ Rev.\  D {\bf 42}, 3488 (1990).



\bibitem{klee}   K.~M.~Lee,
  ``{\it Selfdual nonabelian Chern-Simons solitons},''
  Phys.\ Rev.\ Lett.\  {\bf 66}, 553 (1991);
  ``{\it Relativistic nonAbelian selfdual Chern-Simons systems},''
  Phys.\ Lett.\  B {\bf 255}, 381 (1991).


\bibitem{schap} L.~G.~Aldrovandi and F.~A.~Schaposnik,
  ``{\it Non-Abelian vortices in Chern-Simons theories and their induced effective
  theory},''
  Phys.\ Rev.\  D {\bf 76}, 045010 (2007)
  [arXiv:hep-th/0702209].

\bibitem{zhk}  S.~C.~Zhang, T.~H.~Hansson and S.~Kivelson,
  ``{\it An effective field theory model for the fractional quantum hall effect},''
  Phys.\ Rev.\ Lett.\  {\bf 62} (1988) 82.

\bibitem{pij}  R.~Jackiw and S.~Y.~Pi,
``{\it Soliton Solutions to the Gauged Nonlinear Schrodinger
Equation on the Plane}",  Phys.\ Rev.\ Lett.\  {\bf 64}, 2969
(1990);  ``{\it Selfdual Chern-Simons solitons},''
  Prog.\ Theor.\ Phys.\ Suppl.\  {\bf 107}, 1 (1992).

\bibitem{jip} G.~V.~Dunne, R.~Jackiw, S.~Y.~Pi and C.~A.~Trugenberger,
  ``{\it Selfdual Chern-Simons solitons and two-dimensional nonlinear equations},''
  Phys.\ Rev.\  D {\bf 43}, 1332 (1991)
  [Erratum-ibid.\  D {\bf 45}, 3012 (1992)].

\bibitem{nonrel}  N.~S.~Manton,
  ``{\it First order vortex dynamics},''
  Annals Phys.\  {\bf 256}, 114 (1997)
  [arXiv:hep-th/9701027].


\bibitem{dunne}   G.~V.~Dunne,
  ``{\it Aspects of Chern-Simons theory},''
  arXiv:hep-th/9902115.


\bibitem{tasi}   D.~Tong,
  ``{\it TASI lectures on solitons},''
  arXiv:hep-th/0509216.

\bibitem{moduli}   M.~Eto, Y.~Isozumi, M.~Nitta, K.~Ohashi and N.~Sakai,
  ``{\it Solitons in the Higgs phase: The moduli matrix approach},''
  J.\ Phys.\ A  {\bf 39}, R315 (2006)
  [arXiv:hep-th/0602170].

\bibitem{sy}  M.~Shifman and A.~Yung,
  ``{\it Supersymmetric Solitons and How They Help Us Understand Non-Abelian   Gauge
  Theories},''
  Rev.\ Mod.\ Phys.\  {\bf 79}, 1139 (2007)
  [arXiv:hep-th/0703267].

\bibitem{wang} R. Wang, ``{\it The Existence of Chern-Simons
Vortices}", Commun. Math. Phys. {\bf 137} (1991) 587.


\bibitem{ahiss}  O.~Aharony, A.~Hanany, K.~A.~Intriligator, N.~Seiberg and M.~J.~Strassler,
  ``{\it Aspects of N = 2 supersymmetric gauge theories in three dimensions},''
  Nucl.\ Phys.\  B {\bf 499}, 67 (1997)
  [arXiv:hep-th/9703110].

\bibitem{memirror}  N.~Dorey and D.~Tong,
  ``{\it Mirror symmetry and toric geometry in three dimensional gauge theories},''
  JHEP {\bf 0005}, 018 (2000)
  [arXiv:hep-th/9911094]. \\
 D.~Tong,
  ``{\it Dynamics of N = 2 supersymmetric Chern-Simons theories},''
  JHEP {\bf 0007}, 019 (2000)
  [arXiv:hep-th/0005186].

\bibitem{jr}   R.~Jackiw and P.~Rossi,
  ``{\it Zero Modes Of The Vortex-Fermion System},''
  Nucl.\ Phys.\  B {\bf 190}, 681 (1981).

\bibitem{het}   M.~Edalati and D.~Tong,
  ``{\it Heterotic vortex strings},''
  JHEP {\bf 0705}, 005 (2007)
  [arXiv:hep-th/0703045].



\bibitem{mansch}   N.~S.~Manton and B.~J.~Schroers,
  ``{\it Bundles over moduli spaces and the quantization of BPS monopoles},''
  Annals Phys.\  {\bf 225}, 290 (1993).

\bibitem{wy} A nice description of how index bundles arise in the
context of magnetic monopoles can be found in Appendix A.2 of the
review by  E.~J.~Weinberg and P.~Yi,
  ``{\it Magnetic monopole dynamics, supersymmetry, and duality},''
  Phys.\ Rept.\  {\bf 438}, 65 (2007)
  [arXiv:hep-th/0609055].


\bibitem{vstring}   A.~Hanany and D.~Tong,
  ``{\it Vortex strings and four-dimensional gauge dynamics},''
  JHEP {\bf 0404}, 066 (2004)
  [arXiv:hep-th/0403158].

\bibitem{qhet}     D.~Tong,
  ``{\it The quantum dynamics of heterotic vortex strings},''
  JHEP {\bf 0709}, 022 (2007)
  [arXiv:hep-th/0703235].

\bibitem{syok}  M.~Shifman and A.~Yung,
  ``{\it Non-Abelian semilocal strings in N = 2 supersymmetric QCD},''
  Phys.\ Rev.\  D {\bf 73}, 125012 (2006)
  [arXiv:hep-th/0603134].
  \\ M.~Shifman and A.~Yung,
  ``{\it Heterotic Flux Tubes in N=2 SQCD with N=1 Preserving Deformations},''
  arXiv:0803.0158 [hep-th].


\bibitem{gsy} A.~Gorsky, M.~Shifman and A.~Yung,
  ``{\it Non-Abelian Meissner effect in Yang-Mills
  theories at weak coupling},''
  Phys.\ Rev.\  D {\bf 71}, 045010 (2005)
  [arXiv:hep-th/0412082].

\end{thebibliography}
\end{document}